\documentclass[11pt]{article}
\usepackage{latexsym,graphicx,amssymb,color,amsmath,bbm}
\usepackage{epsfig,lscape,pictex}

\DeclareGraphicsExtensions{.eps}

\def\real{\mathbb{R}}
\def\integer{\mathbb{Z}}

\def\N{\mathbb{N}}
\def\half{\frac{1}{2}} 
\def\t{\tau} 

\begin{document}
\vspace*{-2cm}
\begin{flushright}
q-bio.BM/0512047\\
\end{flushright}

\vspace{0.3cm}

\begin{center}
{\Large {\bf A new series of polyhedra as blueprints for viral capsids \\ in the family of Papovaviridae
 }}\\ 
\vspace{1cm} {\large \bf T.\ Keef\,\footnote{\noindent E-mail: 
{\tt tk506@york.ac.uk}} and
R.\ Twarock\,${}^{1,}$\footnote{\noindent E-mail: 
{\tt rt507@york.ac.uk}}}\\
\vspace{0.3cm} {${}^1$}\em Department of Mathematics \\ University of York\\
\vspace{0.3cm} {${}^3$} Department of Biology \\ University of York \\
York YO10 5DD, U.K. 
\end{center}

\begin{abstract}
In a seminal paper \cite{Caspar:1962} Caspar and Klug established a theory that provides a family of polyhedra as blueprints for the structural organisation of viral capsids. In particular, they encode the locations of the proteins in the shells that encapsulate, and hence provide protection for, the viral genome. 
Despite of its huge success and numerous applications in virology experimental results have provided evidence for the fact  that the theory is too restrictive to describe all known viruses \cite{Casjens:1985}. Especially, the family of Papovaviridae, which contains cancer causing viruses, falls out of the scope of this theory. 

In \cite{Twarock:2004a} we have shown that certain members of the family of Papovaviridae can be described via tilings. In this paper, we develop  a comprehensive mathematical framework for the derivation of {\it all} surface structures of viral particles in this family. We show that this formalism fixes the structure and relative sizes of all particles collectively so that there exists only  one scaling factor that relates the sizes of {\it all} particles  with their biological counterparts. 

The series of polyhedra derived here complements the Caspar-Klug family of polyhedra. It is the first mathematical result that provides a common organisational principle for different types of viral particles in the family of Papovaviridae and paves the way for an understanding of Papovaviridae polymorphism. Moreover, it provides crucial input for the construction of assembly models along the lines of \cite{KTT,KMT}. 
\end{abstract}

\section{Introduction}

Icosahedral symmetry plays a fundamental role for the structure of viruses because it constraints the organisation of the proteins in the viral capsids that protect the viral genome. Based on this observation, Caspar and Klug have developed a landmark theory \cite{Caspar:1962} in which they derive a family of icosadeltahedra, i.e. polyhedra with icosahedral symmetry and faces given by equilateral triangles, that act as blueprints for the organisation of viral capsids. In particular, the corners of the triangular faces of these polyhedra encode the locations of the protein subunits in the capsids. This implies that proteins are organised in clusters of 5 or 6 protein subunits, called {\it pentamers} and {\it hexamers}, respectively.  The polyhedra consist of $20T$ triangular facets, where $T=h^2 + hk + k^2$ with $h\in \N\cup \{0\}$, $k\in \N$ denotes the {\it triangulation number} that parameterises the family of polyhedra. As a consequence, there are precisely 12 pentamers and $10(T-1)$ hexamers in a viral capsid corresponding to the polyhedron of triangulation number $T$ in this family, and the corresponding capsid contains $60T$ protein subunits. 

The family of polyhedra established in Caspar-Klug Theory is fundamental in virology and has a plethora of applications ranging from image analysis of experimental data to the construction of assembly models. 
Despite its huge success experimental evidence has shown that it is too restrictive to account for all known viruses. In particular, viruses in the family of Papovaviridae fall out of the scope of Caspar-Klug Theory and their organisation has therefore been a long-stand open problem in virology \cite{Rayment:1982,Liddington:1991}, which has been formulated by Liddington et al. in 1991 as follows: ``The puzzle is how do the coloured pentamers fit into the hexavalent holes?". In particular, the experiments show that also the protein clusters located off the global 5-fold axes of icosahedral symmetry can be pentamers, which is by construction excluded by the approach of Caspar and Klug because they are working with a hexagonal lattice. 

We have provided a solution to this puzzle based on tiling theory in \cite{Twarock:2004a,Twarock:2005a}, where we have constructed a polyhedron that describes the surface structure of the viral particles observed in \cite{Rayment:1982,Liddington:1991}.
By construction, these models not only encode the locations of the protein subunits but also those of the inter-subunit bonds that connect different pentamers in the capsid. These results have provided the basis for the construction of assembly models for Papovaviridae in \cite{KTT,KMT}, and the tiling approach has moreover paved the way for the study of crosslinking structures \cite{Twarock:2005b}.   

It is the purpose of this work to develop a mathematical framework for the systematic construction of polyhedra associated with viral capsids composed of pentamers throughout, and hence to derive a family of polyhedra that constitute the exceptional cases needed to complement the Caspar-Klug series of polyhedra for the description of viruses.  

Since we are seeking polyhedra that encode the locations of pentamers off the 5-fold axes of the icosahedral group, it is not possible to work with a hexagonal lattice as in the case of the Caspar-Klug construction, and a completely different mathematical approach is hence required. In particular, the coordinates of the vertices have to be retrieved from an icosahedral lattice, i.e. a lattice that is invariant under the action of the icosahedral group. Such a lattice does not exit in three dimensions, but can be inferred from a six-dimensional space via projection. 
This procedure is known from the study of aperiodic structures \cite{Senechal:1996}, which describe the locations of atoms in alloys called quasicrystals \cite{Shechtman:1984}. 
An additional complication arises due to the fact that such projections only lead to discrete point sets if the number of lattice points projected from the higher-dimensional space is restricted in an appropriate way. We use here a restriction that is rooted in the structure of the icosahedral group itself, and can be obtained via a so-called affine extension of that group. It can be used to construct a family of finite dimensional point sets that act as blueprints for the coordinates of the desired polyhedra. 

We obtain three types of polyhedra in this way that correspond to the three species of particles observed in the family of Papovaviridae. The ratios of their radii are completely determined by the mathematical formalism, so that there exists only one scaling parameter that relates the geometries of all polyhedra collectively to the biological setting. It can be used to test the predictions of our theory as we discuss in section \ref{appl}. 

The smallest polyhedron in the series is a triacontahedron, and we therefore call our series of exceptional polyhedra the {\it triacontahedral series}. The 30 faces of the triacontahedron are rhombs. By subdividing each rhomb into two triangles, it can be seen that the triacontahedron is structurally similar to the $T=3$ capsid in Caspar-Klug Theory. However, the coordinates of the vertices are of a different type: while the coordinates of the triacontahedron can be obtained from a higher-dimensional icosahedral lattice via projection, this is not the case for the $T=3$ Caspar-Klug structure. 
Moreover, the vertices of the polyhedra in the triacontahedral series are located on nested spherical shells rather than a  single spherical shell, so that these particles are only approximately spherical. For example, the triacontahedron has 12 vertices (corresponding to an icosahedron) located on one shell  and 20 vertices (corresponding to a dodecahedron) located on a different shell within. 

The polyhedra corresponding to the medium and large sized particles have vertices on three different nested shells. The  vertices at which 5 faces meet mark the centres of the pentamers, and the locations of the protein subunits correspond to the angles of the corresponding faces.   
In combination with the tiling approach cite{Twarock:2004a,Twarock:2005a} it is moreover possible to deduce the bonding structure of the viral capsids they represent as we discuss in section \ref{appl}. 
\medskip

Our mathematical set-up moreover provides a framework for the study of the scaling transformations and rotations that map the coordinates of the polyhedra on other points induced from the higher-dimensional icosahedral lattice by projection. This is important because viral capsids are three- rather than two-dimensional objects, and these transformations can be used to associate a three-dimensional structure with the two-dimensional surfaces of the polyhedron that provides the blueprint for the capsid. This has been demonstrated in \cite{Janner:2005} for the case of Human Rhinovirus. The viral capsid of this virus follows the blueprint of our small shell, the triacontahedron. It has been shown in this reference that the viral capsid is contained between two copies of the triacontahedron that are related by a scaling by the irrational number $\tau:=\half(1+\sqrt{5})$, which corresponds to a transformation of the type we are considering here. 
In section \ref{scale} we show how these transformations can be derived within our formalism for all viruses in the triacontahedral series. This paves the way for a construction of three-dimensional structures associated with  the blueprints provided by the polyhedra for all viruses in this series, and hence to derive the locations also of the proteins that are located within the capsid and delimit the cavity filled by the genetic material. Such information is invaluable for example for the study of scaffold mediated assembly. 
\medskip
 
The paper is organised as follows. In section \ref{proj} we outline the projection formalism that connects a generalised lattice (or quasi-lattice) based on the simple root vectors of the non-crystallographic Coxeter group $H_3$ with the root lattice of $D_6$. In section \ref{nest} we derive finite dimensional subsets of this generalised lattice based on an affine extension of the non-crystallographic Coxeter group $H_3$. In section \ref{const} we use these sets for the construction of the triacontahedral series of polyhedra. The projection picture is extended in section \ref{scale} and a formalism is established that allows us to determine the range of possible scaling transformations, rotations and combinations thereof that map the vertices of the polyhedra on other points in the generalised lattice obtained via projection, because this provides the basis for the construction of three-dimensional models associated with these polyhedra along the lines of \cite{Janner:2005}. A comparison with experimental results is provided in section \ref{appl}. Finally, we conclude with a discussion of the mathematical and biological implications of our results.

\section{Points sets induced from an icosahedral lattices by projection}\label{proj}

We are seeking polyhedra that are symmetric under the icosahedral group. 
Since the icosahedral group does not stabilize a lattice in three dimensions, a 3-dimensional generalised lattice or quasi-lattice has to be inferred from a higher dimensional crystallographic lattice via projection. 

We use the fact that the icosahedral group is crystallographic in 6 dimensions, and construct our quasi-lattice  from  the root lattice of $D_6$ via projection. We remark that it would also have been possible to construct a quasi-lattice in three dimensions via a projection from $\integer^6$ rather than $D_6$ as in \cite{Janner:2005}. However, our approach is more general, because the root lattice of $D_6$ allows for an embedding with maximal symmetry \cite{Baake:1990}.  

Let $\mathbf{e}=\lbrace e_j \vert j=1,\ldots, 6\rbrace$ denote the standard basis in 6 dimensions with $(e_i, e_j)=\delta_{ij}$, $i$, $j=1,\ldots,6$. The root lattice of $D_6$ corresponds to the $\integer$-linear span of the 6 simple root vectors (or roots in short) of the  Coxeter group $D_6$. They can be expressed in terms of the standard basis $\mathbf{e}$ as follows: 
\begin{equation}\label{roots_a}
\begin{array}{rclcrcl}
a_1 & = & e_2-e_1 & \qquad 
a_2 & = & e_1-e_3\\
a_3 & = & e_3-e_6 & \qquad
a_4 & = & e_5+e_6\\
a_5 & = & -e_4-e_5 & \qquad
a_6 & = & e_4-e_5
\end{array}
\end{equation}

We choose a projection to 3 dimensions that maps the basis $\mathbf{e}$ on the vectors pointing to the six non-aligned vertices of an icosahedron. The coordinates of these vectors are given in terms of the 3 dimensional standard basis below, using $\tau=\frac12 (1+\sqrt{5})$: 
\begin{equation}\label{eproject}
\begin{array}{rclcrcl}
e_1 & \mapsto & \frac12 (1,0,\tau) & \qquad
e_2 & \mapsto & \frac12 (\tau,1,0)\\
e_3 & \mapsto & \frac12 (0,\tau,1) & \qquad
e_4 & \mapsto & \frac12 (-1,0,\tau)\\
e_5 & \mapsto & \frac12 (0,-\tau,1) & \qquad
e_6 & \mapsto & \frac12 (\tau,-1,0)\,.
\end{array}
\end{equation}

Hence, by virtue of (\ref{roots_a}) and (\ref{eproject}), the vectors $a_j$, $j=1,\ldots,6$, project on the  vectors in (\ref{aproject}). Coordinates are again expressed in terms of the standard basis in 3 dimensions, with the notation $\tau'=\frac12 (1-\sqrt{5})$ for the Galois conjugate of $\tau$ and the identity $\tau + \tau' =1$:   

\begin{equation}\label{aproject}
\begin{array}{rclcrcl}
a_1 & \mapsto & \widetilde{a_1} =\frac12 (-\tau',1,-\tau) & \qquad
a_2 & \mapsto & \widetilde{a_2} =\frac12 (1,-\tau,-\tau')\\
a_3 & \mapsto & \tau \widetilde{a_3} =\frac12 (-\tau,-\tau^2,1) & \qquad
a_4 & \mapsto & \tau \widetilde{a_2} =\frac12 (\tau, -\tau^2,1)\\
a_5 & \mapsto & \tau \widetilde{a_1} =\frac12 (1,-\tau,-\tau^2) & \qquad
a_6 & \mapsto & \widetilde{a_3} =\frac12 (-1,\tau,-\tau')\,.
\end{array}
\end{equation}
$\widetilde{a_1}$, $\widetilde{a_2}$ and $\widetilde{a_3}$ correspond to the root vectors of the non-crystallographic Coxeter group $H_3$. 

\begin{figure}[h]
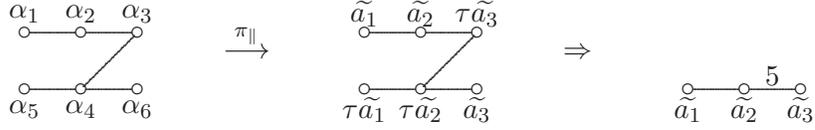

\centerline{
 \beginpicture
 \setcoordinatesystem units <0.75cm, 0.75cm> 
 \setplotarea x from -0.5 to 3.5, y from -.5 to 1.5
 \multiput {$\circ$} at   1 0  2 0  3 0  
                          1 1  2 1  3 1   /
 \plot 1.1 0 1.9 0 /
 \plot 2.1 0 2.9 0 /
 \plot 1.1 1 1.9 1 /
 \plot 2.1 1 2.9 1 /
 \plot 2.95 .95 2.05 0.05 /
 \put {$\alpha_1$} at  1 1.35
 \put {$\alpha_2$} at  2 1.35
 \put {$\alpha_3$} at  3 1.35
 \put {$\alpha_5$} at  1 -.35
 \put {$\alpha_4$} at  2 -.35
 \put {$\alpha_6$} at  3 -.35
 \endpicture
\qquad 
\raisebox{2.5ex}{$ \stackrel{\pi_\|}{\longrightarrow}$}
 \beginpicture
 \setcoordinatesystem units <0.75cm, 0.75cm> 
 \setplotarea x from -0.5 to 3.5, y from -.5 to 1.5
 \multiput {$\circ$} at   1 0  2 0  3 0  
                          1 1  2 1  3 1   /
 \plot 1.1 0 1.9 0 /
 \plot 2.1 0 2.9 0 /
 \plot 1.1 1 1.9 1 /
 \plot 2.1 1 2.9 1 /
 \plot 2.95 .95 2.05 0.05 /
 \put {$\widetilde{a_1}$} at  1 1.35
 \put {$\widetilde{a_2}$} at  2 1.35
 \put {$\tau \widetilde{a_3}$} at  3 1.35
 \put {$\tau \widetilde{a_1}$} at  1 -.35
 \put {$\tau \widetilde{a_2}$} at  2 -.35
 \put {$\widetilde{a_3}$} at  3 -.35
\endpicture
\qquad 
\raisebox{2.5ex}{$\Rightarrow$\qquad}
\beginpicture
 \setcoordinatesystem units <0.75cm, 0.75cm> 
 \setplotarea x from -0.5 to 2.5, y from -.5 to 1.
 \multiput {$\circ$} at 0.0 0   1.0 0  2 0   /
 \plot 0.1 0 0.9 0 /
 \plot 1.1 0 1.9 0 /
 \put {5} at 1.5 0.25
 \put {$\widetilde{a_1}$} at  0 -.35
 \put {$\widetilde{a_2}$} at  1 -.35
 \put {$\widetilde{a_3}$} at  2 -.35
\endpicture}
\caption{Figure illustrating the projection $\pi_{\parallel}$ of the simple roots of $D_6$ on those of $H_3$.}
\label{figProj}
\end{figure}

The projection $\pi_{\parallel}$ in (\ref{aproject}) is illustrated in terms of the Dynkin diagrams of $D_6$ and $H_3$ in Fig. \ref{figProj}. On the left, the Dynkin diagram of $D_6$ is shown, in a folded and hence slightly unconventional form, to illustrate how the simple root vectors of $D_6$ project on the root vectors of $H_3$. In particular, the 5 above the link in the Coxeter diagram of $H_3$ on the right indicates that the angle between the corresponding root vectors is $\pi-\frac\pi 5$, with the 3 over the other link omitted by convention. We remark that this procedure is similar to the projection from $E_8$ on $H_4$ considered in \cite{MoodyPatera}.

The root lattice of $D_6$ is given by $\integer$-linear combinations of the simple roots $a_j$, $j=1,\ldots,6$. Correspondingly, its projection into $\real^3$ via $\pi_{\parallel}$ is given by $\integer[\tau]$-linear combinations of the simple root vectors $\widetilde{a_j}$, $j=1,\ldots,3$, of $H_3$, where $\integer[\tau]$ denotes the extended ring of integers 
\begin{equation}\label{tauint}
\integer[\tau] := \lbrace a+\tau b\, \vert a,b\in\integer\rbrace\,. 
\end{equation} 

Since $\integer[\tau]$ is dense in $\real$, the $\integer[\tau]$-linear combinations of the simple root vectors of $H_3$ are dense in $\real^3$. Thus it is necessary to select subsets thereof which are suitable to guide the search for the vertex sets of the polyhedra. Such sets will be derived in the following section based on an affine extension of $H_3$. 

\section{Nested shell structures via an affine extension of $H_3$}\label{nest}

\begin{figure}[ht]
\begin{center}
\includegraphics[width=3.8cm,keepaspectratio]{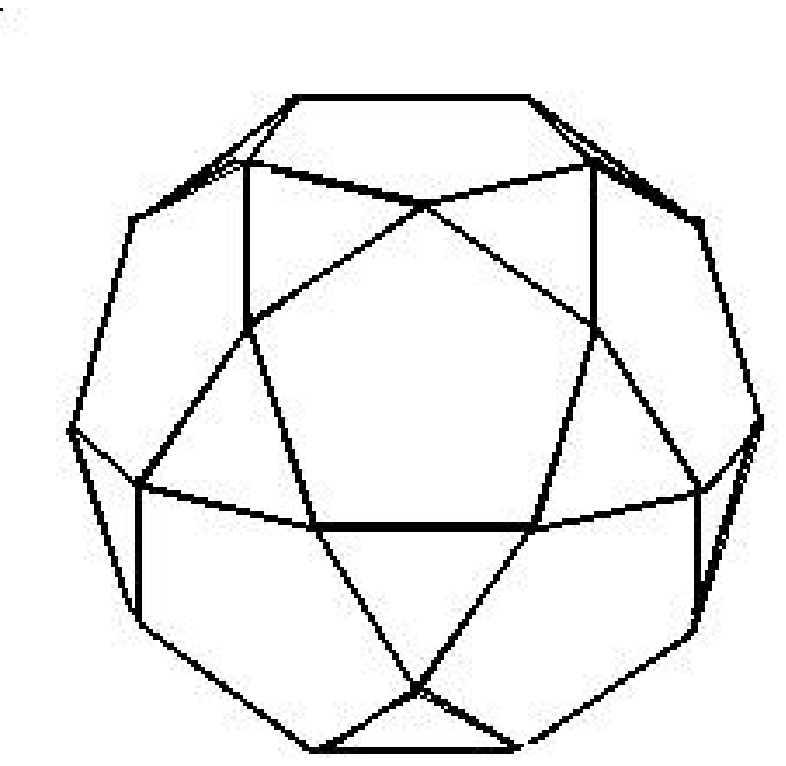}\qquad
\includegraphics[width=3.8cm,keepaspectratio]{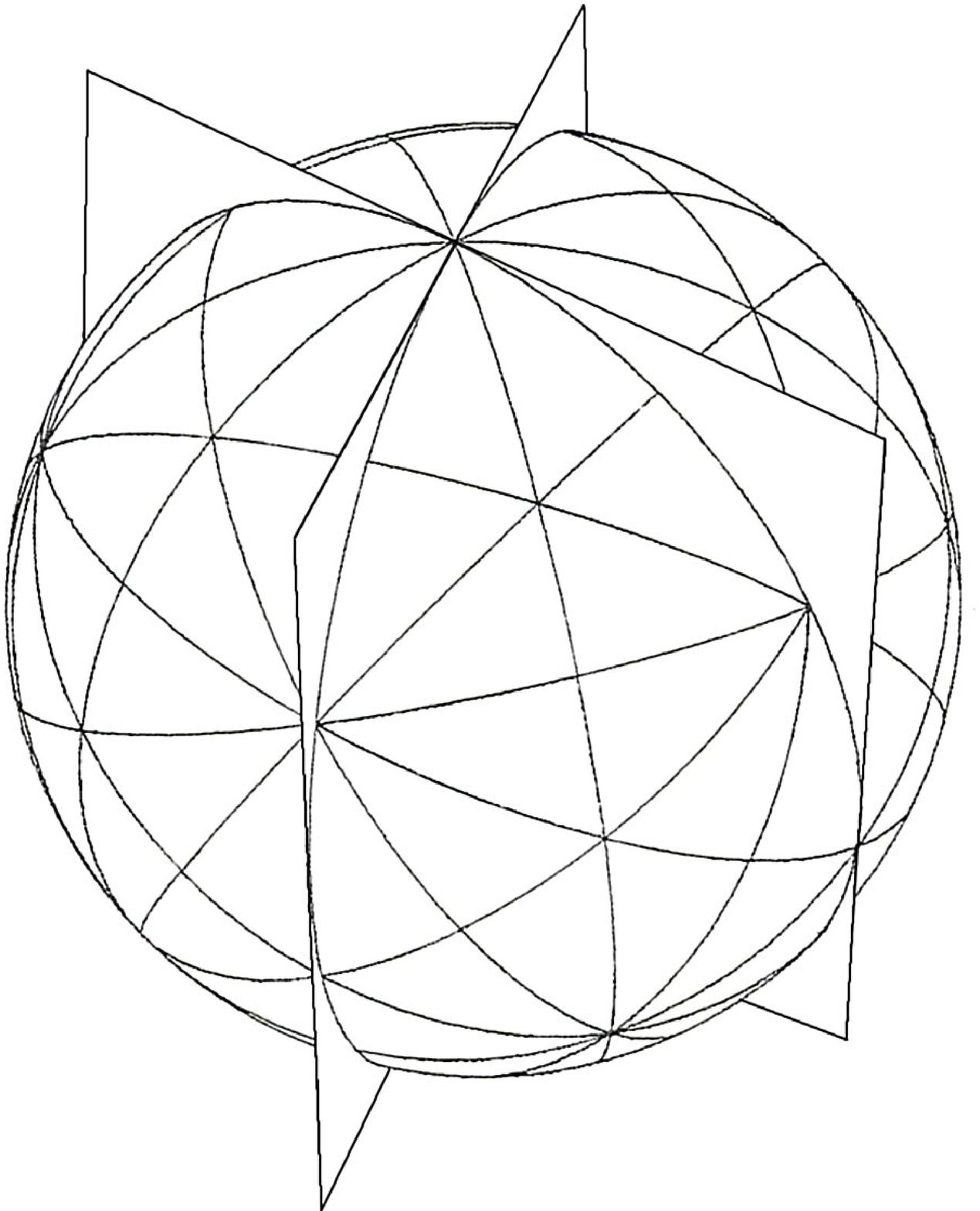}
\end{center}
\caption{The root polytope (left) encoding the locations of the planes of reflection (right).}
\label{SP}
\end{figure}

The simple root vectors $\widetilde{a_j}$, $j=1,\ldots,3$, in (\ref{aproject}) form a basis of the root system of $H_3$ \cite{Humphreys:1992}. 
The complete root system $\Delta$ is given by \cite{CKPS}
\begin{equation}\label{icosH3}
\Delta = \left\lbrace{
\begin{array}{cl}
(\pm 1,0,0) & \mbox{ and all permutations }\\
\half(\pm 1,\pm \t,\pm \tau') & \mbox{ and all even permutations }
\end{array}
}\right\rbrace\,.
\end{equation}
The root vectors in $\Delta$ point to the vertices of an icosidodecahedron. They encode the generators of the Coxeter group $H_3$ as follows. Let $\alpha \in \Delta$ and let $(.\vert .)$ denote the scalar product in $\real^3$. Then 
\begin{equation}
 r_\alpha : x \mapsto x - \frac{2(x\vert\alpha)}{(\alpha\vert\alpha)}
 \quad \mbox{ for } x\in \real^3
\end{equation}
is the Euclidean reflection in the plane orthogonal to the root vector $\alpha$. The icosidodecahedron representing the root system $\Delta$ and two of the reflection planes that are encoded by the root vectors are illustrated in Fig.~\ref{SP}. The intersection of the two planes in the figure corresponds to an axis of 5-fold symmetry,  which intersect with the sphere at two of the 12 5-fold vertices of a (spherical) icosahedron. The other reflection planes are not shown, but their intersections with the surface of the sphere are indicated as geodesics (i.e. as spherical arcs obtained as intersections of planes through the centre of the sphere with the surface of the sphere). The intersections of the geodesics mark the locations of all symmetry axes, and one can hence reconstruct the locations of all 6 five-fold, 15 two-fold and 10 three-fold axes of rotational symmetry of the icosahedral group. 

Note that in contrast to the case of Weyl groups $\frac{2(\alpha\vert\beta)}{(\alpha\vert\alpha)}\in \integer[\tau]$ (instead of $\integer$) for all $\alpha$, $\beta\in\Delta$. Therefore, $\integer$-linear combinations of root vectors in $\Delta$ do not form a crystallographic lattice, and $\Delta$ is therefore called {\it non-crystallographic}. However, it is possible to select subsets which form generalised lattices such as the quasi-lattices known from quasicrystals \cite{Senechal:1996}. 
\medskip

In order to extend the root system in a way compatible with overall icosahedral symmetry the basis of simple root vectors has to be extended. For this, we use the fact that the relations between the simple root vectors are encoded in the Cartan matrix $C$,

\begin{equation}
C:=\left(\frac{2(\alpha_i\mid\alpha_j)}{(\alpha_j|\alpha_j)}\right)_{ij}
= 
\begin{pmatrix}
 2 &    -1 & 0 \cr 
-1 &    2  & -\t \cr
 0 & -\t &   2  
\end{pmatrix}\,.
\end{equation}

We extend this matrix by an additional row and column via a formalism known as affine extensions in the theory of Kac-Moody algebras. The only difference here is that entries stem from the set $\integer[\tau]$ rather than $\integer$. 
We have shown in \cite{Twarock:2002AffH} that the affine extended Cartan matrix of $H_3$ is given by 
\begin{equation}
{\hat C} =
\begin{pmatrix}
 2    &  0  & \t' & 0     \\ 
 0    &  2  &  -1   & 0     \\
\t' &  -1 &   2   & -\t \\
 0    &  0  & -\t &   2  
\end{pmatrix}\,.
\end{equation}
The additional row and column encode a further root vector corresponding to an affine reflection that acts as a translation $T$ by the highest root vector $\alpha_H=
\t\alpha_1+2\t\alpha_2+\t^2\alpha_3$. The three other reflections are cyclic operations of order two and the products of any two of them correspond to rotations around the origin as follows: 
\begin{equation}\label{2reflect}
(r_jr_k)^M= 1 \quad\text{where}\quad
\left\{\begin{matrix}   
M=1 \quad &\text{if}\ &c_{jk}&= 2\\
M=2 \quad &\text{if}\ &c_{jk}&= 0\\
M=3 \quad &\text{if}\ &c_{jk}&=-1\\
M=5 \quad &\text{if}\ &c_{jk}&=-\t\,.
\end{matrix}\right.
\end{equation}
The affine extended group is hence generated by the three reflections $r_1$, $r_2$, $r_3$ and the translation $T$. 
\medskip

It is now possible to construct, based on these operations, finite point sets with icosahedral symmetry  such that all points in the sets have a counterpart in the root lattice of $D_6$. In this sense, they are finite subsets of generalised  lattices induced by projection from $D_6$ and correspond to potential candidates for the locations of the polyhedra we are seeking.  

The point sets are generated iteratively via the action of the generators of the extended group on the origin $0$. If the action of the translation operator $T$ is not restricted, one obtains an infinite point sets that  densely fills $\real^3$ . However, if $T$ acts only a finite number of times, N say, while the action of all other operations is not restricted in order to guarantee icosahedral symmetry of the overall configuration, finite dimensional  point sets ${\mathcal S}(N)$ are obtained which become larger and more dense with increasing $N$.

Due to the fact that $T$ acts as a translation by the highest root vector it has been possible to derive  a simple expression for the point sets ${\mathcal S}(N)$ in terms of the root system $\Delta$ \cite{Twarock:2002Onion}: 
\begin{equation}\label{SN}
{\mathcal S}(N) = \left\lbrace  \sum_{\alpha \in\Delta} n_\alpha \alpha \,\vert \,  n_\alpha \in \N\cup\lbrace 0 \rbrace, \sum_{\alpha \in\Delta} n_\alpha \leq N   \right\rbrace\,.
\end{equation}
In particular, ${\mathcal S}(N)$ consists of all $\N\cup\lbrace 0 \rbrace$-linear combinations of up to $N$ root vectors in $\Delta$. $N$ is called the {\it cut-off parameter} because it limits the number of points in the set. 
\medskip

The sets ${\mathcal S}(N)$ play a crucial role in the following for the construction of the polyhedra.


\section{Construction of the triacontahedral series of polyhedra from $H_3$-induced shell models}\label{const}

By construction, the sets $S(N)$ contain the vertices of the polyhedra in the triacontahedral series. Due to the fact that  we are looking for polyhedra that represent viral capsids given in terms of pentamers that are (with a certain experimental error) equidistant from their neighbouring pentamers, the vertices of the desired polyhedra follow (in a good approximation) the vertices of the Platonic and Archimedean solids, because they correspond to the polyhedra with equidistant edge lengths. They can hence be used as templates for the search of the coordinates of the polyhedra within the sets $S(N)$. Moreover, since we are looking for polyhedra that mark the locations of pentamers, only those Platonic and Archimedean solids are relevant that have 5-coordinated vertices, i.e. vertices at which 5 edges meet. The three solids that fulfil this criterion are the icosahedron, the snub cube and the snub dodecahedron. 

We have computed the sets $S(N)$ up to $N=5$ explicitly. The points are organised on 181 nested shells of radii between 
$R\approx 0.2361$ and $R=5$. $S(5)$ contains all point sets $S(N)$ with $N\leq 5$ by construction, and has been chosen because it is sufficiently dense to provide coordinates for the polyhedra. We have used the icosahedron, the snub cube and the snub dodecahedron, respectively, as a template to search for subsets of $S(5)$ with the desired distribution of points. 
We have thus obtained those coordinates of our polyhedra, that mark the locations of 5-coordinated vertices. We have then determined further vertices (not necessarily on a shell of the same radius) which together with these vertices from the vertex set of the polyhedron. The procedure is discussed explicitly in the following subsections for the individual cases and the coordinates of the vertices corresponding to centres of pentamers are provided in the Appendix.

\subsection{The small species}

\begin{figure}[ht]
\begin{center}
\includegraphics[width=3.5cm,keepaspectratio]{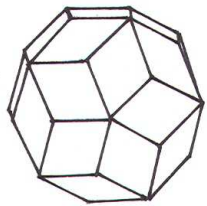}
\end{center}
\caption{The triacontahedron corresponding to the polyhedron of the small species.}
\label{triac}
\end{figure}

The vertex set of the icosahedron occurs for the first time in the set $S(3)$ on a shell of radius $R\approx 1.1756$. 
There are two ways of obtaining polyhedra that match this vertex set. One of them is the icosahedron itself and corresponds to the start of the Caspar-Klug series. The other one is the rhombic triacontahedron with 30 rhombic faces.  
The remaining vertices of this polyhedron correspond to the vertices of a dodecahedron located on shell of a radius $R\approx 1.0705$ (i.e. there is  a scaling factor of about 1.098 between the shells that contain the two different types of vertices). The triacontahedron is shown in Fig.~\ref{triac}. 

\begin{figure}[ht]
\begin{center}
\includegraphics[width=5cm,keepaspectratio]{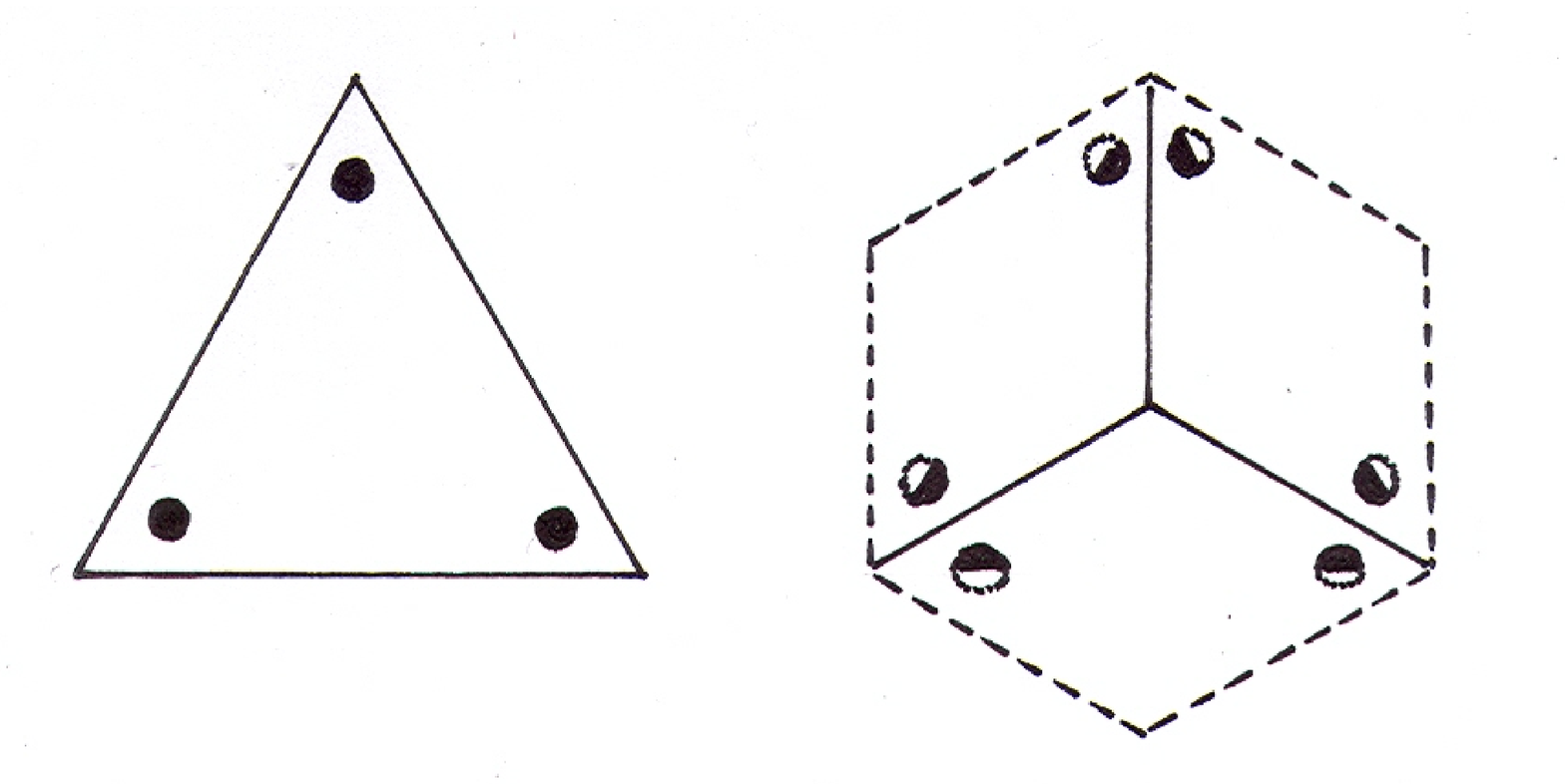}
\end{center}
\caption{The locations of proteins on the triangular faces of the icosahedron (left) are rotated by an angle of $\frac{2\pi}{10}$ with respect to the locations of the proteins on the rhombs  (right).}
\label{rotate}
\end{figure}

Since only those vertices at which 5 faces meet mark the locations of pentamers, both cases correspond to capsids with 12 pentamers located on the axes of 5-fold symmetry of the icosahedral group. However, only the centres of the pentamers coincide in both cases, and the {\it orientations} of the pentamers are {\it different}: because the proteins are located in the corners of the faces (shown schematically as dots in Fig.~\ref{rotate}) they differ by a rotation by an angle of $\frac{2\pi}{10}$ as illustrated in this figure. It shows a face of the icosahedron with three protein subunits on the left versus three faces of the triacontahedron on the right with 2 protein subunits each. A superposition of both figures shows that the respective proteins are rotated with respect to each other. 

This difference has far-reaching consequences for the structure of the capsids. In particular, in one of the cases crosslinking structures are possible while they cannot occur for geometrical reasons in the other case, as has been demonstrated in \cite{Twarock:2005b}. 

The case of the triacontahedron is hence essentially different from the $T=1$ case in the Caspar-Klug family and marks the start of a different series of polyhedra.

\subsection{The intermediate species}

\begin{figure}[ht]
\begin{center}
\includegraphics[width=3.5cm,keepaspectratio]{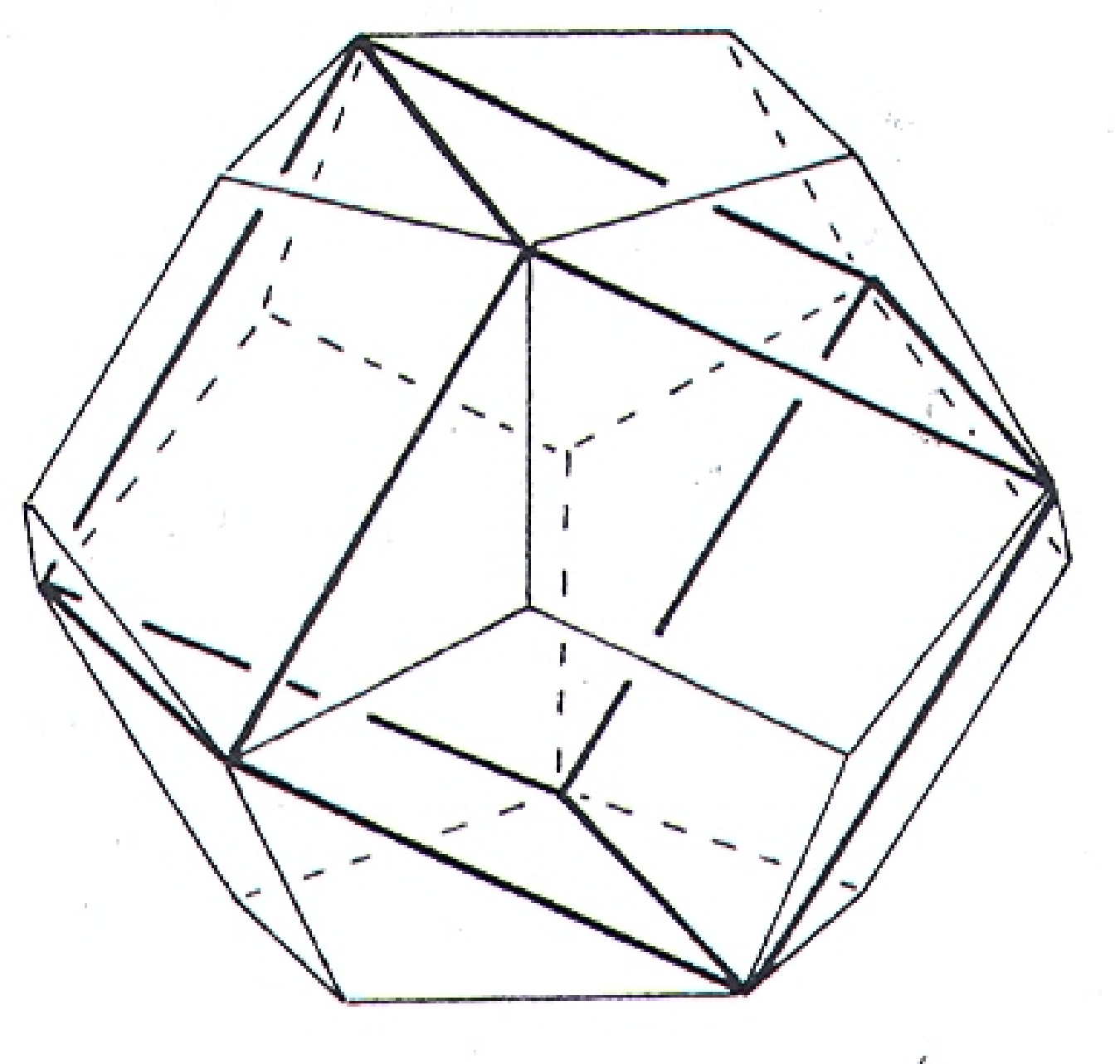}\qquad
\includegraphics[width=3.5cm,keepaspectratio]{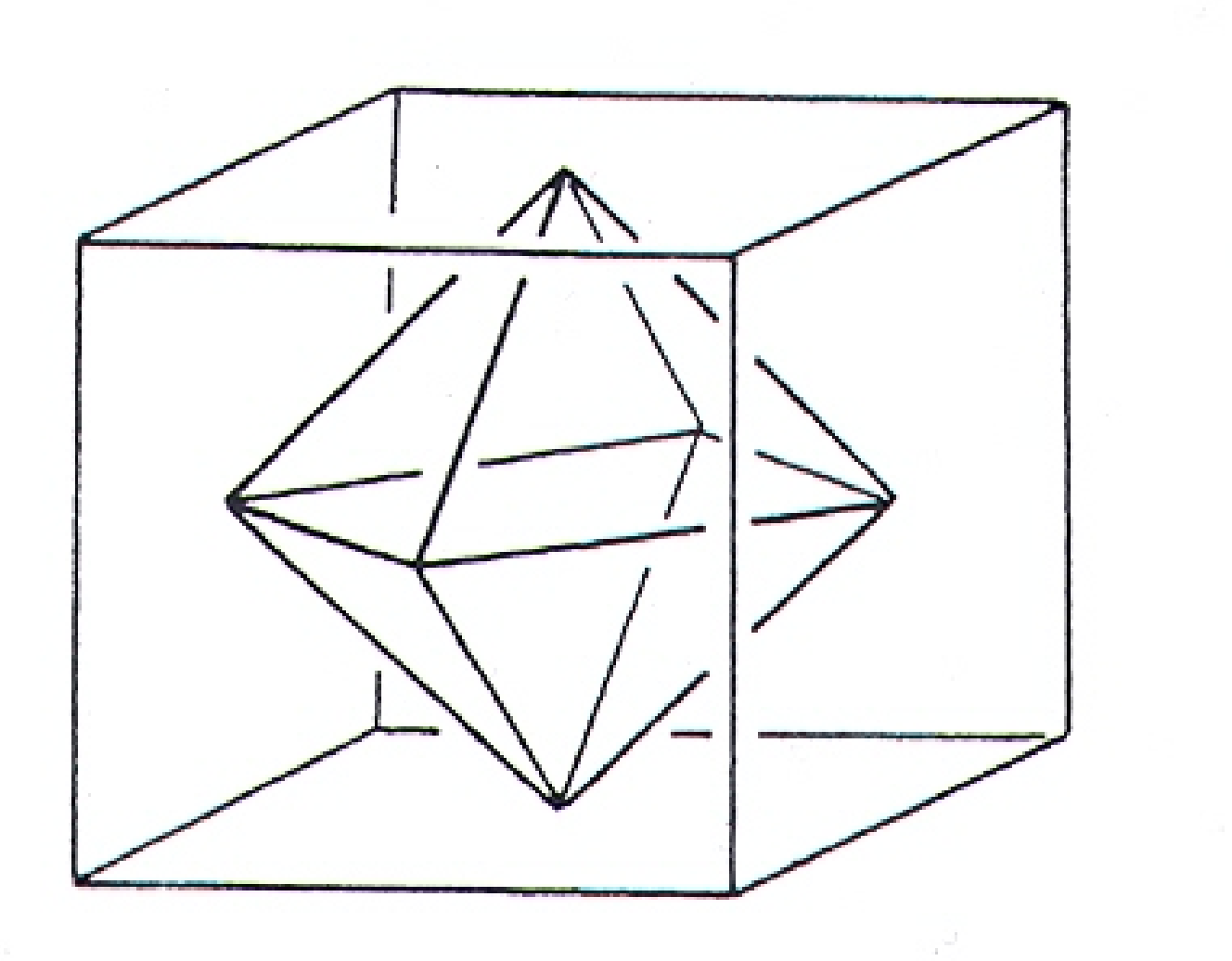}
\end{center}
\caption{A cube inscribed into a dodecahedron (left) and an octahedron into a cube (right).}
\label{inclusions}
\end{figure}

Due to the fact that the snub cube has octahedral rather than icosahedral symmetry, we need to use the geometric inclusions illustrated in Fig.~\ref{inclusions} in order to locate its vertices within the point set $S(5)$.  

The points of $S(5)$ that are closest to the vertex set of the snub cube occur on a shell of radius $R\approx 2,3199$, which appears in $S(4)$ for the first time (i.e. it is not contained in a set $S(N)$ with $N<4$). The edges are not equidistant in the case of the snub cube as expected, but the deviations are small. In particular, one obtains the following values for the three different distances $A$, $B$ and $C$ between neighbouring points in the set. 
\begin{equation}\label{ABC}
A \approx 1,7481\,, \quad B \approx 1,6625\,, \quad C\approx 1,8783\,.
\end{equation}

\begin{figure}[ht]
\begin{center}
\includegraphics[width=6.5cm,keepaspectratio]{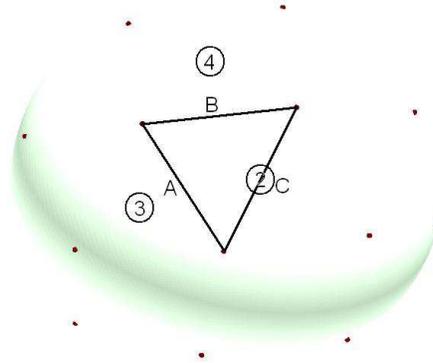}
\end{center}
\caption{The triangle $\Delta_{ABC}$ in the octahedral case.}
\label{octtri}
\end{figure}

The point set is illustrated in Fig.~\ref{octtri} (using the 3d-Grapher software) in relation to the locations of the 2-, 3- and 4-fold symmetry axes of the octahedral group. Moreover, the triangle $\Delta_{ABC}$ formed from the distances $A$, $B$ and $C$ has been superimposed.

From the triangle $\Delta_{ABC}$ the polyhedron is constructed via the following argument. The line denoted as $C$ is centred on a global two-fold axis of the symmetry group. The face containing that edge is hence 2-fold symmetric about that axis. We compute the intersection of this axis with the edge $C$, and determine a line that goes through that intersection point, is perpendicular to $C$ and intersects the sphere in two points equidistant from the intersection point. We then determine the intersection of this line with the line through the global 4-fold axis (marked as an encircled 4 in Fig.~\ref{octtri}) and the centre of edge $B$. The intersection is located within the sector of the sphere that is characterised by the triangle $\Delta_{ABC}$ and determines one further type of vertex of the polyhedron that does not correspond to the centre of a pentamer. A third type of vertex is located on the global 3-fold axis, that is indicated as an  encircled 3 in Fig.~\ref{octtri}. There is flexibility in the construction in that the vertices on the global 3- and 4-fold axes do not need to be located on a shell of the same radius as the vertices representing pentamers. However, due to the fact that they are not marking locations of pentamers, their exact location is not important for our purpose. It is important to note though that a variation of the radii of the shells on which they are located causes changes in the angles of the faces. 

\begin{figure}[ht]
\begin{center}
\includegraphics[width=6cm,keepaspectratio]{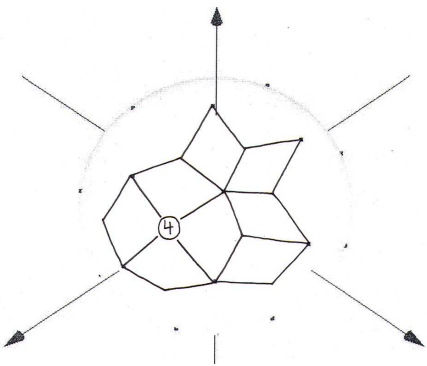}
\end{center}
\caption{Schematic representation of the polyhedron corresponding to the medium species.}
\label{tiling1}
\end{figure}

One hence obtains the polyhedron illustrated schematically in Fig.~\ref{tiling1}. The fundamental domain of the octahedral group corresponds to the triangle between the 2-fold, 3-fold, and 4-fold axes marked as encircled 2s, 3s and 4s in Fig.~\ref{octtri} and the entire polyhedron can be obtained from it via the action of the octahedral group. Therefore, only a part of the polyhedron overlapping with the fundamental domain  is shown,  because the rest of the polyhedron is implied by symmetry. It is a schematic representation in that the vertices not marking the locations of the pentamers have been drawn on the same shell. However, this illustration shows all features of the polyhedron that are needed to read off the locations of the protein subunits, which are located in those corners of the faces that meet in multiples of 5 at a vertex.

It is a polyhedron in terms of rhomb- and kite-shaped faces. The coordinates of the vertices that mark the centres of the pentamers are given in the Appendix. We remark that the angles of the tiles around these vertices are not equal, but the deviations are small. In particular, the angles on the kites are larger than those on the rhombs. This fact may account for the experimental observation that viral particles corresponding to such a polyhedron have only been obtained in vitro, but have never been observed in vivo. 

 



\subsection{The large species}

The construction of the polyhedron corresponding to the large species is analogous to the procedure used before. In this case, we use the vertex set of the snub dodecahedron as a template. The subset of $S(5)$ closest to it is located on a shell of radius $R\approx 3,1247$ and appears in $S(5)$ for the first time. 
\begin{figure}[ht]
\begin{center}
\includegraphics[width=6cm,keepaspectratio]{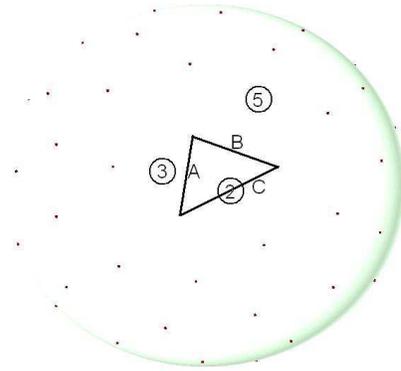}
\end{center}
\caption{The triangle $\Delta_{ABC}$ in the octahedral case.}
\label{7tri}
\end{figure}

The three different distances between neighbouring points are given in (\ref{ABC3}). 
\begin{equation}\label{ABC3}
A \approx 1,3708\,, \quad B \approx 1,7520\,, \quad C\approx 1,4364\,.
\end{equation}

\begin{figure}[ht]
\begin{center}
\includegraphics[width=6cm,keepaspectratio]{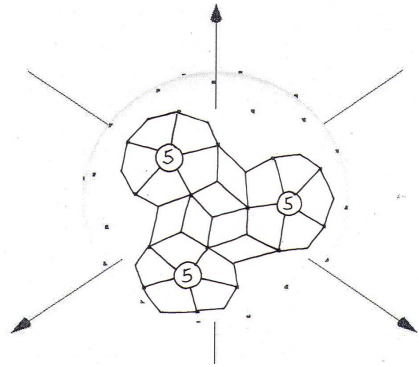}
\end{center}
\caption{Schematic representation of the polyhedron corresponding to the large species.}
\label{tiling2}
\end{figure}

They define  the triangle $\Delta_{ABC}$ in  Fig.~\ref{7tri}. 
We start again by fixing the rhomb on the 2-fold axis containing C, as well as the angles of the kite-shaped faces around the 5-fold axes. This determines the vertex within the sector corresponding to the triangle $\Delta_{ABC}$ as before. The radius of the shell that determines the locations of the global 5-fold vertices can be computed such that the angles on the kite-shaped faces at the vertices meeting rhombs match those of the rhombs containing C. Finally, the angles of the rhombs around the global 3-fold axes can be adjusted by varying the radius of the shell on which they are located. In this way, the polyhedron in Fig.~\ref{tiling2} is obtained. As before, it is a schematic representation that encodes all important information about the polyhedron. The fundamental domain of the icosahedral group corresponds to the triangle between the 2-fold, 3-fold, and 5-fold axes marked as encircled 2s, 3s and 5s in Fig.~\ref{7tri} and the entire polyhedron can be obtained from it via the action of the icosahedral group. It is hence sufficient to represent only part of the polyhedron,  provided that this contains the fundamental domain, in order to specify the polyhedron completely. 

The coordinates of the vertices that mark the locations of the pentamers are given in the Appendix. They correspond to 12 vertices located on the 5-fold axes of icosahedral symmetry, as well as further 60 vertices off these axes. 
\bigskip

Finally, we remark that the set $S(5)$ contains the vertices of all three types of polyhedra. Hence, the geometrical formalism fixes the sizes of the three different species of particles in relation to each other, so that there is indeed only one free parameter that relates the overall mathematical structure, i.e. the small, medium and large polyhedron collectively, to the biological setting. We determine the corresponding scaling factor in section \ref{appl}. 


\section{Scaling transformations and rotations}\label{scale}

In order to determine the range of possible scaling transformations, rotations and combinations thereof that map the vertices of the polyhedra onto other points in a set $S(N)$ (possibly with different $N$) one needs to extend the projection picture presented in section \ref{proj}. 

\subsection{The extended projection picture} 

Under the action of the icosahedral group, $\real^6$ decomposes into two irreducible orthogonal subspaces, which we  denote as $E_\perp$ and $E_\parallel$. This decomposition induces a projection of the root lattice of $D_6$ onto two different copies of the root system of $H_3$, which are located in $E_\perp$ and $E_\parallel$, respectively. 

\begin{figure}[h]
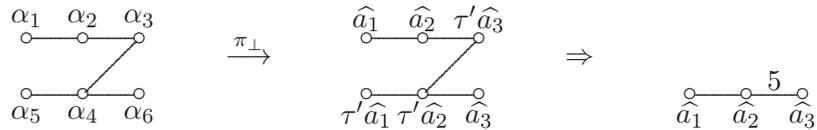

\centerline{
 \beginpicture
 \setcoordinatesystem units <0.75cm, 0.75cm> 
 \setplotarea x from -0.5 to 3.5, y from -.5 to 1.5
 \multiput {$\circ$} at   1 0  2 0  3 0  
                          1 1  2 1  3 1   /
 \plot 1.1 0 1.9 0 /
 \plot 2.1 0 2.9 0 /
 \plot 1.1 1 1.9 1 /
 \plot 2.1 1 2.9 1 /
 \plot 2.95 .95 2.05 0.05 /
 \put {$\alpha_1$} at  1 1.35
 \put {$\alpha_2$} at  2 1.35
 \put {$\alpha_3$} at  3 1.35
 \put {$\alpha_5$} at  1 -.35
 \put {$\alpha_4$} at  2 -.35
 \put {$\alpha_6$} at  3 -.35
 \endpicture
\qquad 
\raisebox{2.5ex}{$ \stackrel{\pi_{\perp}}{\longrightarrow}$}
 \beginpicture
 \setcoordinatesystem units <0.75cm, 0.75cm> 
 \setplotarea x from -0.5 to 3.5, y from -.5 to 1.5
 \multiput {$\circ$} at   1 0  2 0  3 0  
                          1 1  2 1  3 1   /
 \plot 1.1 0 1.9 0 /
 \plot 2.1 0 2.9 0 /
 \plot 1.1 1 1.9 1 /
 \plot 2.1 1 2.9 1 /
 \plot 2.95 .95 2.05 0.05 /
 \put {$\widehat{a_1}$} at  1 1.35
 \put {$\widehat{a_2}$} at  2 1.35
 \put {$\tau' \widehat{a_3}$} at  3 1.35
 \put {$\tau' \widehat{a_1}$} at  1 -.35
 \put {$\tau' \widehat{a_2}$} at  2 -.35
 \put {$\widehat{a_3}$} at  3 -.35
\endpicture
\qquad 
\raisebox{2.5ex}{$\Rightarrow$\qquad}
\beginpicture
 \setcoordinatesystem units <0.75cm, 0.75cm> 
 \setplotarea x from -0.5 to 2.5, y from -.5 to 1.
 \multiput {$\circ$} at 0.0 0   1.0 0  2 0   /
 \plot 0.1 0 0.9 0 /
 \plot 1.1 0 1.9 0 /
 \put {5} at 1.5 0.25
 \put {$\widehat{a_1}$} at  0 -.35
 \put {$\widehat{a_2}$} at  1 -.35
 \put {$\widehat{a_3}$} at  2 -.35
\endpicture}
\caption{Figure illustrating the projection $\pi_{\perp}$ of the simple roots of $D_6$ on those of $H_3$.}
\label{figProj2}
\end{figure}

The projection of the simple roots of $D_6$ onto $E_\parallel$ has been considered in Fig.~\ref{figProj} in section \ref{proj}. Analogously, we obtain a second projection, denoted as $\pi_{\perp}$, which acts as shown in Fig.~\ref{figProj2}.

The coordinates of the vectors in $E_{\perp}$ are given in (\ref{atilde2}). 
\begin{equation}\label{atilde2}
 \begin{array}{rcl}
 \widehat{a_1} & = & \frac12 (-\tau, -\tau',  1), \\
 \widehat{a_2} & = & \frac12 (1, -\tau, -\tau'), \\
 \widehat{a_3} & = & \frac12 (-1, -\tau,  \tau'), \\
 \end{array} \quad
 \begin{array}{rcl}
 \tau' \widehat{a_1} & = & \frac12 (1, -{\tau'}^2, -\tau'), \\
 \tau' \widehat{a_2} & = & \frac12 (\tau', 1, -{\tau'}^2), \\
 \tau' \widehat{a_3} & = & \frac12 (-\tau', 1,  {\tau'}^2). \\
 \end{array}
\end{equation}

\begin{figure}[ht]
\begin{center}
\includegraphics[width=7cm,keepaspectratio]{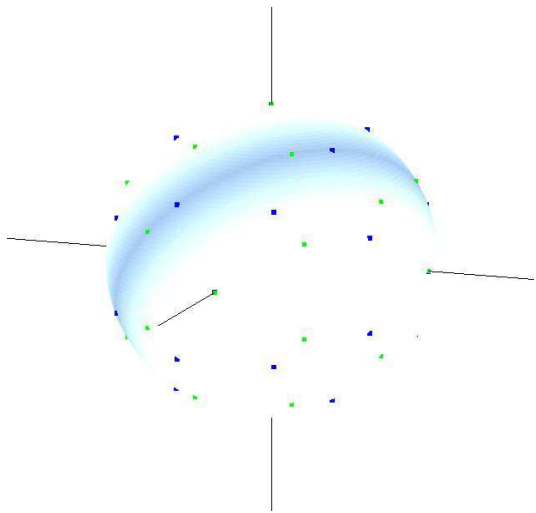}
\end{center}
\caption{The root systems of the two $H_3$ copies on $E_\parallel$ and $E_\perp$.}
\label{PN}
\end{figure}

A comparison of (\ref{atilde2}) and (\ref{aproject}) shows that the copies of $H_3$ in $E_{\parallel}$ and $E_{\perp}$ are related by an interchange of $\tau$ and $\tau'$ in the coordinates of the root vectors. 
Hence, each simple root vector $\widetilde{a_j}$, $j=1,\ldots,3$, of the $H_3$-copy in $E_\parallel$ has a counterpart $\widehat{a_j}$, $j=1,\ldots,3$, in one of the simple root vectors of the $H_3$-copy in $E_\perp$ and vice versa, related by an exchange of $\tau$ and $\tau'$ in their coordinates. This applies also to the root system $\Delta$ in (\ref{icosH3}). The root systems of the two $H_3$-copies in $E_\parallel$ and $E_\perp$ are shown superimposed onto each other in Fig.~\ref{PN} as dark and light dots, respectively, to demonstrate the geometric implications of this relation. 

As mentioned in section \ref{proj}, the root lattice of $D_6$ is given by $\integer$-linear combinations of the simple roots $a_j$, $j=1,\ldots,6$, and its projection $\pi_\parallel(D_6)$ onto $E_\parallel$ corresponds to the $\integer[\tau]$-linear combinations of the simple root vectors $\widetilde{a_j}$, where $\integer[\tau]$ is the extended ring of integers defined in (\ref{tauint}). It is a dense set in $\real^3$ due to the properties of $\integer[\tau]$. 
Hence, a formalism is required to select a discrete subset of $\pi_\parallel(D_6)$. 

This is possible via the cut-and-project method known from the study of quasicrystal \cite{Senechal:1996}. In particular, we  
define an automorphism $^*$ as a mapping from $\pi_{\parallel}(D_6)$ to $\pi_{\perp}(D_6)$ that acts on a vector $x\in \pi_{\parallel}(D_6)$ by exchanging $\tau$ and $\tau'$ in all coordinates. For any connected,  bounded area $\Omega\subset E_\perp$ we define point sets $\Sigma(\Omega)$ by 
\begin{equation}\label{cutpro}
 \Sigma(\Omega) = 
 \left\lbrace x \in \pi_\parallel(D_6) \Big| x^{*} \in\Omega
 \right\rbrace \subset E_\parallel\,. 
 \end{equation}
The construction is illustrated in (\ref{projscheme}).
\begin{equation}\label{projscheme}
\begin{array}{ccccc}
 \Omega\subset E_\perp &  \stackrel{\pi_\perp}{\longleftarrow} & \real^6  & 
  \stackrel{\pi_\parallel}{\longrightarrow} & \Sigma(\Omega)\subset E_\parallel \\
 & & \cup & & \\
 & & D_6 & & 
\end{array}
\end{equation}
The point sets $\Sigma(\Omega)$ have been extensively studied in the quasicrystal literature, where they are used to model the locations of atoms in alloys \cite{Shechtman:1984,Senechal:1996}.

\subsection{Scaling transformations and rotations via the projection picture}

The sets $S(N)$ in (\ref{SN}) are by construction finite dimensional subsets of the cut-and-project quasicrystals $\Sigma(\Omega)$ in (\ref{cutpro}), i.e.~one has the set inclusion $S(N)\subset \Sigma(\Omega)$ for suitably chosen areas $\Omega$. A canonical choice for $\Omega$ is the Voronoi domain of the root lattice of $D_6$ projected to $E_\perp$ \cite{Papadopolos:1999}. It corresponds to a rhombic triacontahedron centred about the origin with edge length $1/\sqrt{2}$ and inradius $R\approx 5.32$. 

The coordinates of the vertices of our polyhedra that mark the locations of pentamers are provided in the Appendix. They  can be used to check that this choice of $\Omega$ is large enough to contain their counterparts in $E_{\perp}$ under the mapping ${}^*$. Let $\mathcal{V}_j$, 
$j\in\{1,2,3\}$, denote these vertex sets for the polyhedra corresponding the large,  medium and  small particle, respectively, and let $\mathcal{V}_j^*:=\{  x^* \,\vert \,x\in \mathcal{V}_j\}$ be their counterpart in $E_{\perp}$. Then one obtains the following radii for the shells in $E_{\perp}$ on which the sets 
$\mathcal{V}_j^*$ are located: 
\begin{equation}
\begin{array}{rcl}
\mathcal{V}_1^* & : & R\approx 3.77\,, \quad R\approx 4.16\\
\mathcal{V}_2^* & : & R\approx 2.76\\
\mathcal{V}_3^* & : & R\approx 1.9\,.
\end{array}
\end{equation}
Since these are all smaller that the inradius $R\approx 5.32$ of the triacontahedron that corresponds to $\Omega$, the vertex sets of the polyhedra are indeed contained in $\Sigma(\Omega)$. 

Following \cite{Janner:2005} we call a transformation {\it crystallographic} if it can be expressed as an $N$-dimensional matrix with integral entries in an $N$-dimensional vector space. All transformations that map the vertex sets of our polyhedra on points in $\Sigma(\Omega)$ are by construction crystallographic in this sense. 
It is hence possible to determine the corresponding scaling transformations, rotations and combinations thereof via the projection picture. In particular, they can be studied as transformations in $E_\perp$ and correspond to those transformations that map $\mathcal{V}_j^*$ onto points in $\pi_\parallel(D_6)\subset \Omega$. 

\begin{itemize}
\item {\bf Scaling transformations:}

All transformations that act as contractions in $\Omega$, that is scalings by $z=z_1+\tau z_2 \in \integer [\tau]$ (cf. (\ref{tauint})) with modulus $\vert z \vert <1$  correspond to crystallographic scaling transformations. In particular, they map $\mathcal{V}_j^*$ on a set ${\mathcal{V}_j^*}_z:= \{  zx^* \,\vert \,x^*\in \mathcal{V}_j^*\}$. Using ${z^*}^*=z$  this induces a  scaling of $\mathcal{V}_j$  by 
$z^* \in E_{\parallel}$, which acts as a stretching transformation since $\vert z^* \vert >1$. 

Hence, every $z\in \integer [\tau]$ with $\vert z \vert <1$ induces a stretching transformation $T_{z^*}$ on 
$\mathcal{V}_j$ in $E_{\parallel}$. An example is $z=1-\tau = \tau'$ which induces a scaling of the index set $\mathcal{V}_j$ by $\tau$.  

\item {\bf Rotations:}

Rotations can again be studied as rotations in $\Omega$. Since the shells on which the sets $\mathcal{V}_j^*$ are located are  entirely contained within the inradius of the triacontahedron that defines $\Omega$, any rotation in $\Omega$ that maps $\mathcal{V}_j^*$ on points in $\pi_{\perp}(D_6) \cap \Omega$ again induces a transformation of $\mathcal{V}_j$ in 
$\pi_{\parallel}(D_6)$.  

\end{itemize}

The same holds for combinations of rotations and scaling transformations. In particular, also the polygrammal transformations that play an important role in \cite{Janner:2005} can easily be studied  via the projection formalism in this way.




\section{Validation against experimental results}\label{appl}

Our approach can be tested based on the following two criteria: the predictions for the ratios of the radii of the different species of particles which are fixed completely by our mathematical formalism, as well as the predictions concerning the locations of the inter-subunit bonds in the capsids that are implied by the structure of the polyhedra. 

\subsection{Predictions concerning the relative radii of particles in different species}

Our construction fixes the radii of the different polyhedra with respect to each other, and there is hence only one free parameter, $\alpha$ say, that relates the geometries of all particles collectively with their biological counterparts. In particular, we obtain a radius of $R_L=3.1247$ for the large particle, of $R_M=2.3199$ for the medium particle and of $R_S=1.1756$ (using the radius at the global five-fold vertices where the pentamers are located) for the small particle. The ratios of these are  
\begin{equation}\label{ratios}
\frac{R_{M}}{R_S} \approx 1.9734\,, \qquad \frac{R_L}{R_S} \approx 2.658 \,.
\end{equation}

The polymorphic assemblies of the major capsid protein of Simian Virus 40, which is a member of the family of Papovaviridae, has been studied in \cite{Kanesashi:2003}. They report the occurrence of three different types of particles: a large particle with a diameter of approximately 40 to 45 nm, a medium sized particle with a diameter of size 25 to 35 nm and a small particles with a diameter of approximately 20 nm. It has not been possible in that work to distinguish whether the medium sized particles correspond to the octahedral particles consisting of 24 pentamers reported in \cite{Salunke:1989}. The radii of these particles are, taking averages, approximately 
\begin{equation}\label{ratiosApprox}
R_L^K \approx 21.25 \mbox{nm} \,, \qquad  R_M^K \approx 15 \mbox{nm} \,, \mbox{ and }\qquad R_S^K \approx 10\mbox{nm} \,.
\end{equation}

Since these are approximate values, we test whether any combination with small deviations from these values is compatible with the ratios obtained theoretically in (\ref{ratios}). One finds that the combination 
\begin{equation}\label{ratiosApprox2}
R_L^K =21.26 \mbox{nm} \,, \qquad  R_M^K =15.79 \mbox{nm} \,, \mbox{ and }\qquad R_S^K = 8\mbox{nm} 
\end{equation}
corresponds to the ratios in (\ref{ratios}). The correspondence with the values in (\ref{ratiosApprox}) is remarkably close. It implies that the ratio between large and medium size particles are in excellent agreement with the experimental results. Moreover, given that the exact value of the small sized  shell is more difficult to determine experimentally  than those of the larger ones, a deviation of  $2$nm from the experimentally measured average can be considered as an equally good agreement with experimental results. 

Based on (\ref{ratiosApprox2}), it is possible to compute the scaling factor, that relates the geometry of the shells of radii $R_L$, $R_{M}$ and $R_S$ in the theoretical model collectively with the biological system, as   
\begin{equation}\label{scalefactor}
\alpha=\frac{R_\lambda^K}{R_\lambda}\approx 6,80\mbox{nm}\, \mbox{ with } \lambda\in\{S,M,L\}\,.
\end{equation}

\subsection{Predictions concerning the bonding structure}

\begin{figure}[ht]
\begin{center}
\includegraphics[width=5cm,keepaspectratio]{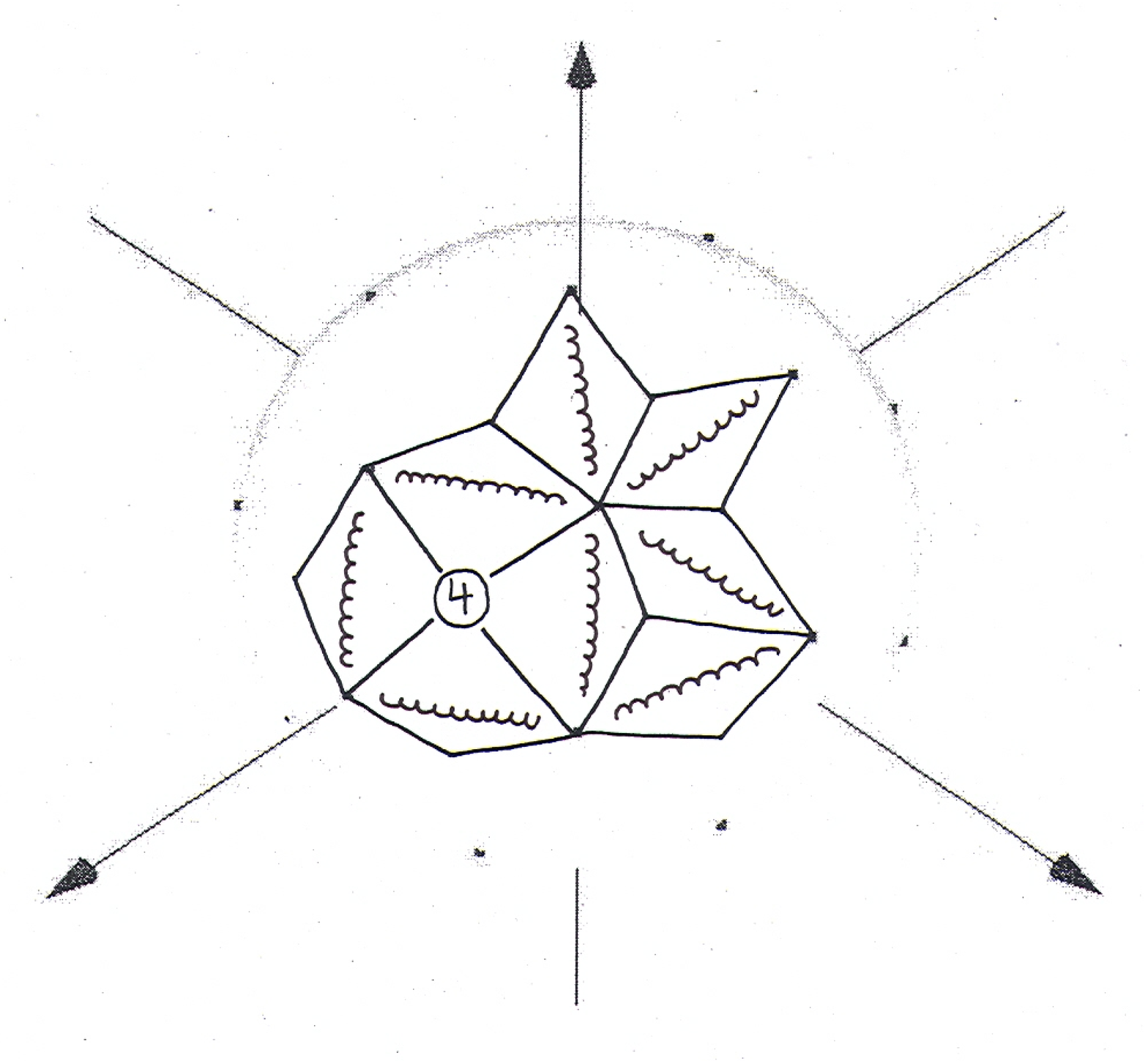}\qquad
\includegraphics[width=5cm,keepaspectratio]{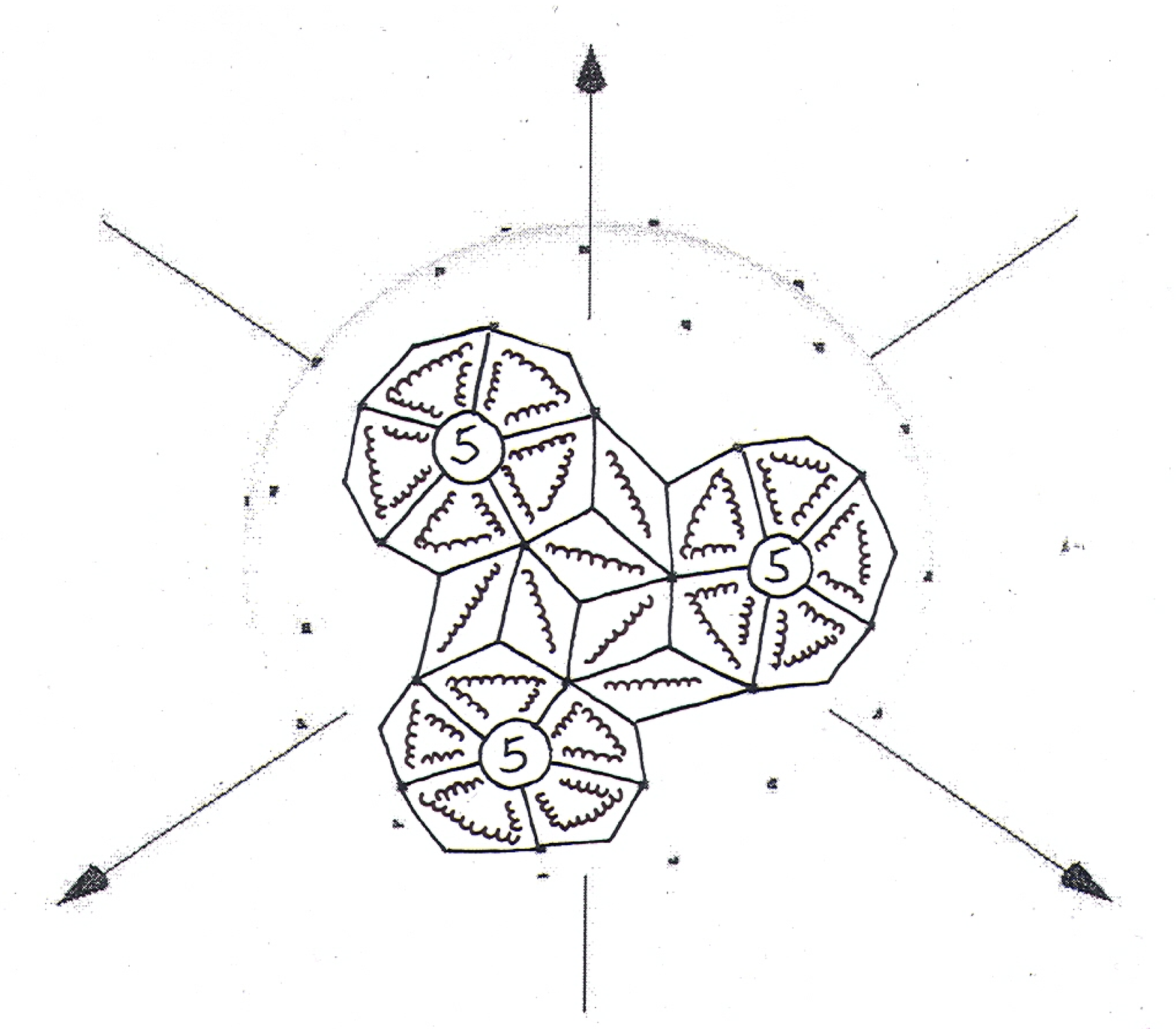}
\end{center}
\caption{The bonding structure of medium and large particles implied by the tiling approach. }
\label{bondingstructure}
\end{figure}

The polyhedra provide information on the locations of intersubunit bonds between the proteins in different pentamers in the capsid. Based on the tiling approach developed earlier \cite{Twarock:2004a,Twarock:2005a,Twarock:2005b}, the faces of the polyhedra encode the locations of these interactions as follows: a face with three corners corresponding to centres of  pentamers in the capsid represents a trimer interaction between the proteins located in these corner; respectively, a face with two such corners  represents a dimer interaction between these two protein subunits. The bonding structure implied by the polyhedra of the medium and large species are illustrated in Fig.~\ref{bondingstructure}, where the locations of the bonds are shown schematically by spiral arms. 

The combination of dimer and trimer bonds predicted by our theory for the large shell in Fig.~\ref{bondingstructure} on the right coincides with the experimental results for the locations of the C-terminal arm extensions observed in \cite{Modis:2002}. It would be interesting to validate also the predictions for the bonding structure of the other species  experimentally.  

\section{Conclusions}

The results in this paper provide answers to the two open mathematical questions stated in the conclusion of \cite{Janner:2005}. 

Firstly, we have addressed the question of how the series of polyhedra that starts with the triacontahedron and contains   all polyhedra corresponding to all-pentamer configurations, can be completed. For this, we have derived a mathematical method that combines a projection of the root lattice of $D_6$ with an affine extension of the non-crystallographic Coxeter group $H_3$ in order to determine finite subsets of generalised lattices that encode the vertices of these polyhedra. In this way, we have been able to establish the triacontahedral series that complements the family of polyhedra in Caspar-Klug Theory. 

Secondly, the mathematical formalism developed here provides a framework for the systematic analysis of crystallographic scaling transformations and rotations for polyhedra with vertices in the triacontahedral series. These are important as they can be used to associate three-dimensional structures with the blueprints provided by the surface structures of the  polyhedra as demonstrated in \cite{Janner:2005} for the case of Human Rhinovirus, which corresponds to the small particle in our tricontahedral series. In this respect, our results pave the way for an  analysis of the interior organisation of all viral capsids corresponding to the triacontahedral series.   
\medskip

From the biological point of view, our triacontahedral series closes an important gap because it complements the series of Caspar-Klug polyhedra. Similar as the in the case of the Caspar-Klug series, this paves the way for a broad spectrum of  applications in virology. In particular, due to the fact that Papovaviridae contain cancer-causing viruses, a better understanding of their structure is highly desirable to assist the development of new anti-viral drugs. For example, based on the structure of the polyhedra it is possible to develop assembly models that explain how viral capsids self-assemble from their capsid proteins. The theory developed here provides a framework for the generalisation of earlier work \cite{KTT,KMT} to the case of simultaneous assembly of different species of particles. Such models are important in order to study how viral assembly can be misdirected as a means to interfere with the viral replication cycle.

\section*{Acknowledgements}
RT has been supported by an EPSRC Advanced Research Fellowship. TK has been supported by the EPSRC grant GR/T26979/01.


\section*{Appendix}

In the Appendix, we provide the coordinates of the vertices of the polyhedra that correspond to the centres of pentamers. 

The vertices corresponding to the 60 pentamers  off the 5-fold axes of the large particles are located on a shell 
of radius $R\approx 3.1247$. Their coordinates are given in (\ref{global1}). 

\begin{equation}\label{global1}
\begin{array}{ll}
 \displaystyle\frac12(2\tau-{\tau}',1-2{\tau}',6-\tau) &
  \displaystyle\frac12(2\tau-{\tau}',-(1-2{\tau}'),-(6-\tau)) \\
   \displaystyle\frac12(-(2 \tau-{\tau}'),1-2 {\tau}',6-\tau) &
   \displaystyle\frac12(-(2 \tau-{\tau}'),1-2 {\tau}',-(6-\tau)) \\
   \displaystyle\frac12(1-2 {\tau}',6-\tau,2 \tau-{\tau}') &
   \displaystyle\frac12(1-2 {\tau}',-(6-\tau),-(2 \tau-{\tau}')) \\
   \displaystyle\frac12(-(1-2 {\tau}'),-(6-\tau),2 \tau-{\tau}') &
   \displaystyle\frac12(-(1-2 {\tau}'),6-\tau,-(2 \tau-{\tau}')) \\
   \displaystyle\frac12(6-\tau,2 \tau-{\tau}',1-2 {\tau}') &
   \displaystyle\frac12(6-\tau,-(2 \tau-{\tau}'),-(1-2 {\tau}')) \\
   \displaystyle\frac12(-(6-\tau),-(2 \tau-{\tau}'),1-2 {\tau}') &
   \displaystyle\frac12(-(6-\tau),2 \tau-{\tau}',-(1-2 {\tau}')) \\
   \displaystyle\frac12({\tau}'+4\tau,-1-4{\tau}',\tau) &
   \displaystyle\frac12({\tau}'+4\tau,-(-1-4{\tau}'),-\tau) \\
   \displaystyle\frac12(-({\tau}'+4\tau),-(-1-4{\tau}'),\tau) &
   \displaystyle\frac12(-({\tau}'+4\tau),-1-4{\tau}',-\tau) \\
   \displaystyle\frac12(-1-4{\tau}',\tau,{\tau}'+4\tau) &
   \displaystyle\frac12(-1-4{\tau}',-\tau,-({\tau}'+4\tau)) \\
   \displaystyle\frac12(-(-1-4{\tau}'),-\tau,{\tau}'+4\tau) &
   \displaystyle\frac12(-(-1-4{\tau}'),\tau,-({\tau}'+4\tau)) \\
   \displaystyle\frac12(\tau,{\tau}'+4\tau,-1-4{\tau}') &
   \displaystyle\frac12(\tau,-({\tau}'+4\tau),-(-1-4{\tau}')) \\
   \displaystyle\frac12(-\tau,-({\tau}'+4\tau),-1-4{\tau}') &
   \displaystyle\frac12(-\tau,{\tau}'+4\tau,-(-1-4{\tau}')) \\
   \displaystyle\frac12(\tau-3{\tau}',1+2{\tau}',2+\tau) &
   \displaystyle\frac12(\tau-3{\tau}',-(1+2{\tau}'),-(2+\tau)) \\
   \displaystyle\frac12(-(\tau-3{\tau}'),-(1+2{\tau}'),2+\tau) &
   \displaystyle\frac12(-(\tau-3{\tau}'),1+2{\tau}',-(2+\tau)) \\
   \displaystyle\frac12(-(1+2{\tau}'),-(2+\tau),\tau-3{\tau}') &
   \displaystyle\frac12(-(1+2{\tau}'),2+\tau,-(\tau-3{\tau}')) \\
   \displaystyle\frac12(1+2{\tau}',2+\tau,\tau-3{\tau}') &
   \displaystyle\frac12(1+2{\tau}',-(2+\tau),-(\tau-3{\tau}')) \\
   \displaystyle\frac12(2+\tau,-(\tau-3{\tau}'),-(1+2{\tau}')) &
   \displaystyle\frac12(2+\tau,\tau-3{\tau}',1+2{\tau}') \\
   \displaystyle\frac12(-(2+\tau),\tau-3{\tau}',-(1+2{\tau}')) &
   \displaystyle\frac12(-(2+\tau),-(\tau-3{\tau}'),1+2{\tau}') \\
   \displaystyle\frac12({\tau}'+2+2\tau,-(2\tau-2{\tau}'-1),-(-2-\tau-2{\tau}')) &
   \displaystyle\frac12({\tau}'+2+2\tau,2\tau-2{\tau}'-1,-2-\tau-2{\tau}') \\
   \displaystyle\frac12(-({\tau}'+2+2\tau),2\tau-2{\tau}'-1,-(-2-\tau-2{\tau}')) &
   \displaystyle\frac12(-({\tau}'+2+2\tau),-(2\tau-2{\tau}'-1),-2-\tau-2{\tau}') \\
   \displaystyle\frac12(2\tau-2{\tau}'-1,-2-\tau-2{\tau}',{\tau}'+2+2\tau) &
   \displaystyle\frac12(2\tau-2{\tau}'-1,-(-2-\tau-2{\tau}'),-({\tau}'+2+2\tau)) \\
   \displaystyle\frac12(-(2\tau-2{\tau}'-1),-(-2-\tau-2{\tau}'),{\tau}'+2+2\tau) &
   \displaystyle\frac12(-(2\tau-2{\tau}'-1),-2-\tau-2{\tau}',-({\tau}'+2+2\tau)) \\
   \displaystyle\frac12(-(-2-\tau-2{\tau}'),-({\tau}'+2+2\tau),2\tau-2{\tau}'-1) &
   \displaystyle\frac12(-(-2-\tau-2{\tau}'),{\tau}'+2+2\tau,-(2\tau-2{\tau}'-1)) \\
   \displaystyle\frac12(-2-\tau-2{\tau}',{\tau}'+2+2\tau,2\tau-2{\tau}'-1) &
   \displaystyle\frac12(-2-\tau-2{\tau}',-({\tau}'+2+2\tau),-(2\tau-2{\tau}'-1)) \\
 \displaystyle (3,{\tau}',-{\tau}') &
 \displaystyle (3,-{\tau}',{\tau}') \\
 \displaystyle (-3,-{\tau}',{\tau}') &
 \displaystyle (-3,{\tau}',-{\tau}') \\
 \displaystyle (-{\tau}',{\tau}',3) &
 \displaystyle (-{\tau}',-{\tau}',-3) \\
 \displaystyle ({\tau}',-{\tau}',3) &
 \displaystyle ({\tau}',{\tau}',-3) \\
 \displaystyle (-{\tau}',-3,-{\tau}') &
 \displaystyle (-{\tau}',3,{\tau}') \\
 \displaystyle ({\tau}',3,-{\tau}') &
 \displaystyle ({\tau}',-3,{\tau}') 
\end{array}
\end{equation}

\newpage 

\begin{figure}[ht]
\begin{center}
\includegraphics[width=9cm,keepaspectratio]{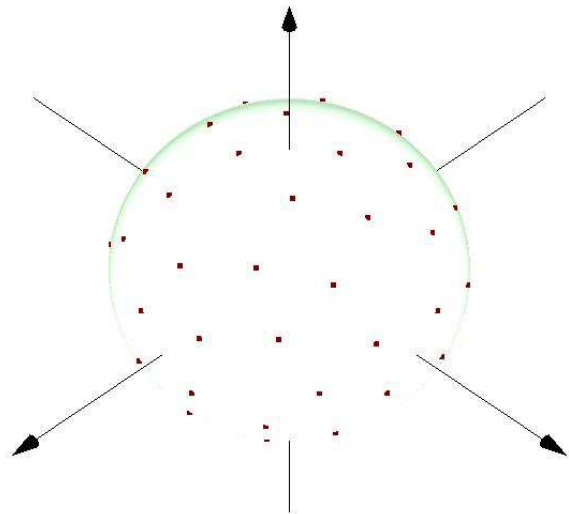}
\end{center}
\caption{The 60 vertices in (\ref{global1}) corresponding to the centres of the pentamers off the global symmetry axes.}
\label{T7}
\end{figure}

The locations of the coordinates in (\ref{global1}) on a sphere are illustrated in Fig.~\ref{T7} via the 3d-Grapher software. 

The vertices corresponding to the centres of the remaining pentamers of the large particles point to the vertices of an icosahedron inscribed into a sphere of radius $R= 5 \times 0.7265 \approx 3.6325$. The coordinates are given in (\ref{coor2}).

\begin{equation}\label{coor2}
 \displaystyle\frac12(\pm(\tau-3-{\tau}'),0,\pm 2{\tau}')  \mbox{ and all cyclic permutations}\,.
\end{equation}
\bigskip





 The vertices of the polyhedron corresponding to the centres of the pentamers in the capsids of the medium sized particles  are located on a sphere of radius $R \approx 2.3199$. Their coordinates are given in (\ref{coor3}). 

\begin{equation}\label{coor3}
\begin{array}{ll}
 (-1-\tau+2{\tau}',1,-3-\tau)/2  &
 (-1,3+\tau,-1-\tau+2{\tau}')/2 \\
 (-3-\tau,1+\tau-2{\tau}',-1)/2 &
 (3{\tau}'-1-\tau,2{\tau}',0)/2 \\
 (-2{\tau}',0,3{\tau}'-1-\tau)/2 &
 (0,-3{\tau}'+1+\tau,-2{\tau}')/2 \\
 ({\tau}'-2,2{\tau}'-2\tau+1,-\tau)/2 &
 (-2{\tau}'+2\tau+1,\tau,{\tau}'-2)/2 \\
 (\tau,\tau,2-\tau) &
 (\tau,-\tau,2-\tau) \\
 (\tau,-2+\tau,\tau) &
 (-\tau,-2+\tau,\tau) \\
 (2-\tau,-\tau,\tau) &
 (2-\tau,-\tau,-\tau) \\
 (2+\tau,3{\tau}',2-\tau)/2 &
 (-3{\tau}',-2+\tau,2+\tau)/2 \\
 (3{\tau}',-2+\tau,2+\tau)/2 &
 (2-\tau,-2-\tau,-3{\tau}')/2 \\
 (1+{\tau}',-3-\tau,2-\tau+{\tau}')/2 & 
 (3+\tau,-2+\tau-{\tau}',1+{\tau}')/2 \\
 (2-\tau+{\tau}',-1-{\tau}',3+\tau)/2 &
 (-{\tau}',2,-1) \\
 (-2,1,-{\tau}') &
 (-1,{\tau}',-2) \,.
\end{array}
\end{equation}

\begin{figure}[ht]
\begin{center}
\includegraphics[width=9cm,keepaspectratio]{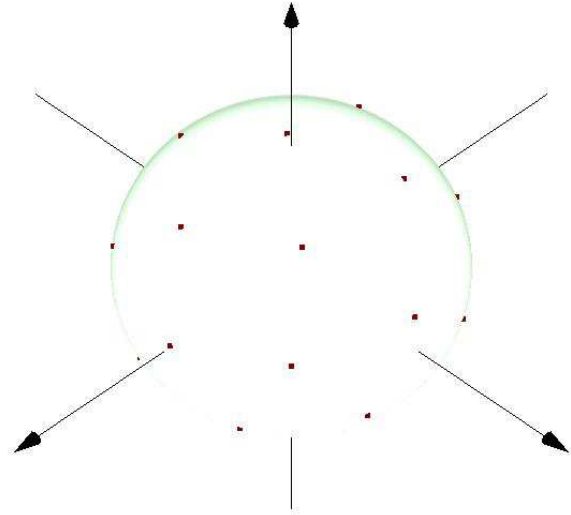}
\end{center}
\caption{The 24 vertices in (\ref{coor3}) corresponding to the centres of the pentamers.}
\label{Oct2.3199}
\end{figure}

The locations of the coordinates in (\ref{coor3}) on a sphere are illustrated in Fig.~\ref{Oct2.3199} via the 3d-Grapher software. 

\medskip

 The vertices of the triacontahedron that correspond to the centres of the pentamers of the small particles are located on a sphere of radius $R \approx 1.1756$ with coordinates given in (\ref{coor4}). 

\begin{equation}\label{coor4}
(\pm 1, 0, \pm {\tau}')  \mbox{ and all cyclic permutations}\,.
\end{equation}

\end{document}